\documentclass[pra,showpacs,twocolumn]{revtex4}%
\usepackage{amsmath}
\usepackage{amsfonts}
\usepackage{amssymb}
\usepackage{graphicx}

\providecommand{\U}[1]{\protect\rule{.1in}{.1in}}

\begin{document}
\title{Quadrature interferometry for nonequilibrium ultracold atoms in optical lattices}
\author{E. Tiesinga}
\affiliation{Joint Quantum Institute, National Institute of Standards and Technology and
University of Maryland, 100 Bureau Drive, Stop 8423 Gaithersburg, Maryland
20899-8423, USA}
\author{P. R. Johnson}
\affiliation{Department of Physics, American University, Washington DC 20016, USA}

\pacs{37.10.Jk, 37.25.+k, 67.85.Hj, 06.20.Dk}
\date{\today}

\begin{abstract}
We develop an interferometric technique for making time-resolved measurements
of field-quadrature operators for nonequilibrium ultracold bosons in optical
lattices. The technique exploits the internal state structure of magnetic
atoms to create two subsystems of atoms in different spin states and lattice
sites. A Feshbach resonance turns off atom-atom interactions in one spin
subsystem, making it a well-characterized reference state, while atoms in the
other subsystem undergo nonequilibrium dynamics for a variable hold time.
Interfering the subsystems via a second beam-splitting operation,
time-resolved quadrature measurements on the interacting atoms are obtained by
detecting relative spin populations. The technique can provide quadrature
measurements for a variety of Hamiltonians and lattice geometries (e.g.,
cubic, honeycomb, superlattices), including systems with tunneling, spin-orbit
couplings using artificial gauge fields, and higher-band effects. Analyzing
the special case of a deep lattice with negligible tunneling, we obtain the
time evolution of both quadrature observables and their fluctuations. As a
second application, we show that the interferometer can be used to measure
atom-atom interaction strengths with super-Heisenberg scaling $\bar{n}^{-3/2}$
in the mean number of atoms per lattice site, and standard quantum limit
scaling $M^{-1/2}$ in the number of lattice sites. In our analysis, we require
$M\gg1\ $and for realistic systems $\bar{n}$ is small, and therefore the
scaling in total atom number $N=\bar{n}M$ is below the Heisenberg limit;
nevertheless, measurements testing the scaling behaviors for interaction-based
quantum metrologies should be possible in this system.

\end{abstract}
\maketitle

\section{Introduction}

Ultracold atoms in optical lattices \cite{Lewenstein2007,Jaksch2005,Bloch2012}
are versatile systems for studying nonequilibrium physics
\cite{Greiner2002,Schutzhold2006,Fischer2008,Pollack2010,Simon2011,Lin2011,Daley2012,Hung2012}%
, particularly for making time-resolved measurements of system observables.
For example, lattice collapse-and-revival experiments
\cite{Greiner2002,Will2010,Anderlini2006,Sebby2007} follow atom populations in
the $k=0$ quasimomentum of nonequilibrium Bose gases versus system evolution
time. Our goal is to extend
the power of these experiments to enable time-resolved measurements of a
greater variety of system observables and Hamiltonians. A second goal is to
use nonequilibrium collapse-and-revival dynamics to investigate the physics of
interaction-based (or nonlinear) quantum metrology, which exploits multibody
dynamics to characterize a system and its parameters
\cite{Boixo2007,Boixo2008,Choi2008,Liu2010,Grond2011,Napolitano2011,Giovannetti2011,Javanainen2012,Benatti2011}%
.

In this paper, we design an interaction-based interferometer for making
time-resolved measurements of quadrature operators \cite{Walls2008} of
nonequilibrium matter fields in optical lattices. Measurement of quantum-field
quadratures $X_{k}=(e^{-i\zeta_{k}}A_{k}+e^{i\zeta_{k}}A_{k}^{\dagger}%
)/2,$where $A_{k}$ annihilates atoms with quasimomentum $k$ and spin state
$a$, and $\zeta_{k}$ is a tunable phase, can provide direct determination of
the order parameter for a Bose gas \cite{Kitagawa2011}, as opposed to
inferring it from density measurements \cite{Pethick2008,Pitaevskii2003}. The
technique, analogous to homodyne detection, works as follows. Exploiting the
internal state structure of ultracold atoms, we split a superfluid into two
subsystems---the arms of the interferometer---consisting of atoms in different
spin states and lattice sites. A Feshbach resonance \cite{Chin2010} turns off
atom-atom interactions in one spin subsystem, making it a well-characterized
reference \textquotedblleft beam,\textquotedblright\ while atoms in the other
subsystem undergo nonequilibrium dynamics for a variable hold time.
Interfering the subsystems via a second beam-splitting operation, spatial
quadrature observables are obtained by measuring spin populations.

As a second, distinct application of the interferometer, we show that
atom-atom interaction strengths can be determined from the frequencies of
matter-wave collapse-and-revival oscillations with \textquotedblleft
super-Heisenberg\textquotedblright%
\ \cite{Boixo2007,Boixo2008,Choi2008,Liu2010}\ scaling $\bar{n}^{-3/2}$ in the
mean number of atoms per lattice site $\bar{n},$ and standard quantum limit
(SQL) scaling $M^{1/2}$ in the number of lattice sites $M$. In contrast, we
find that the optimal scaling from conventional collapse-and-revival dynamics,
which do not exploit entanglement of motional and spin degrees of freedom, is
$\bar{n}^{-3/4}.$ Our analysis requires $M\gg1$ and for realistic lattice
systems $\bar{n}\lesssim5;$ consequently, the scaling in total atom number
$N=M\bar{n}$ is below the Heisenberg bound. Nevertheless, we find that
measurements testing the predicted scaling behaviors for interaction-based
metrologies should be possible in a lattice system using quadrature
measurements. We note that measuring the quadratures of nonequilibrium lattice
fields, our first goal, is less experimentally demanding than determining
interaction strengths with optimal precision. Recently, super-Heisenberg
scaling has been realized in another system \cite{Napolitano2011}.

In the following, we first describe the interferometric technique and propose
implementations. We next analyze the special case of atoms held in a deep,
cubic lattice with negligible tunneling, and obtain analytic predictions for
the time evolution of both quadrature observables and their fluctuations. We
emphasize, however, that the technique can provide time-resolved quadrature
measurements for more complicated Hamiltonians and lattice geometries (e.g.,
cubic, honeycomb, superlattices
\cite{Sebby2006,Vaucher2008,Barmettler2008,Assam2010}), including systems with
tunneling, spin-orbit couplings using artificial gauge fields \cite{Lin2011},
and higher-band effects \cite{Muller2007,Bakr2010,Soltan2012}, and a method
for studying quench dynamics \cite{Zurek2005,Cucchietti2007,Chen2011}. Related
theoretical developments on collapse-and-revival physics can be found in
\cite{Buchhold2011,Fischer2012,Gramsch2012,Daley2012}.

\section{Interferometry technique}

\begin{figure}[ptb]
\begin{center}
\includegraphics[scale=1.0,trim=0 0 0 0,angle=0,clip]{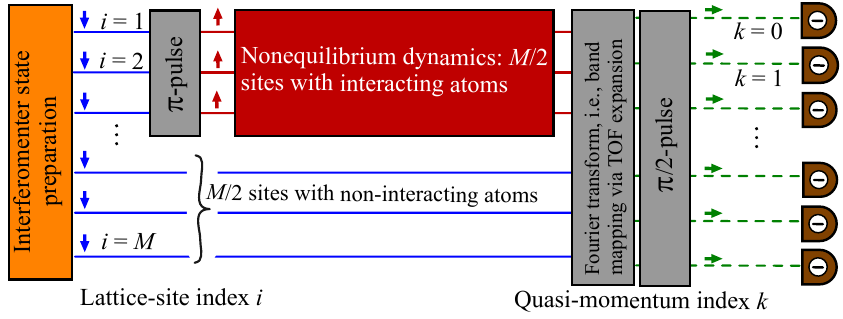} \vspace*{-28pt}
\end{center}
\caption{(Color online) Schematic of quadrature operator interferometer for
two-state atoms held in $M$ optical lattice sites. Time evolution is left to
right. The left-most box represents initial state preparation. Solid and
dashed lines correspond to atoms in lattice sites $i$ and Bloch states with
quasimomenta $k$, respectively. Bold-red and thin-blue lines are atoms in spin
states $|a\rangle$ and $|b\rangle$, respectively. Green lines indicate
superposition states. Arrows specify spin states. Gray and red boxes
correspond to unitary operations described in the text. Detectors (brown) at
right measure population differences between spin states at quasimomenta $k$.
}%
\label{sketch}%
\end{figure}

Figure~\ref{sketch} shows a schematic of the quadrature interferometer. First,
a superfluid of two-component bosonic atoms, with spin states $\left\vert
a\right\rangle $ and $\left\vert b\right\rangle ,$ are prepared in an optical
lattice with all atoms in state $\left\vert b\right\rangle $. Interactions
between $b$-state atoms ($b$ atoms) are then slowly turned off with a Feshbach
resonance, such that the initial interferometer state is a coherent state with
$k=0$ quasimomentum in the lowest Bloch band of an optical lattice potential.
We therefore have $|\psi_{k=0}\rangle=\exp(\gamma B_{k=0}^{\dagger}%
-\gamma^{\ast}B_{k=0})|0\rangle$, where $|0\rangle$ is the empty lattice,
$B_{k}$ annihilates $b$ atoms with quasimomentum $k$, and the mean total atom
number is $\bar{N}=|\gamma|^{2}$. We can re-express this as the separable
state $|\psi_{k=0}\rangle=\prod_{i=1}^{M}\exp(\beta b_{i}^{\dagger}%
-\beta^{\ast}b_{i})|0\rangle$, where $B_{k=0}=\sum_{i}b_{i}/\sqrt{M}$ and
$b_{i}$ annihilate $b$ atoms in the lowest Wannier function of lattice site
$i$. The mean atom number per lattice site is ${\bar{n}}=|\beta|^{2}=\bar
{N}/M$. For notational simplicity, we suppress the band index and vectorial
nature of $i$ and $k$. As an aside, assuming an initial Fock state
$(B_{k=0}^{\dag})^{N}\left\vert 0\right\rangle /\sqrt{N!}$ (e.g., see
\cite{Schachenmayer2011}) leads to the same interferometer results found below
with finite-size corrections of order $M^{-3/2}$.

The lattice depth is next increased to turn off tunneling without exciting
atoms to higher bands, and the Wannier function can be approximated by the
energetically lowest local orbital in a site. At this point atoms neither
tunnel nor interact. A $\pi$ pulse between internal states $a$ and $b$ is then
applied to half of the lattice sites, transforming the system state to
$\prod_{i=1}^{M/2}\exp(\beta a_{i}^{\dagger}-\beta^{\ast}a_{i})\otimes
\prod_{j=M/2+1}^{M}\exp(\beta b_{j}^{\dagger}-\beta^{\ast}b_{j})|0\rangle$,
where $a_{i}$ annihilates $a$ state atoms ($a$ atoms) in the lowest orbital of
site $i$. The $\pi$ pulse is the first beam splitter of the interferometer,
creating two arms with $a$ and $b$ atoms, respectively.

After the $\pi$ pulse, atoms are held for a variable evolution time $t$. The
$a$ atoms interact and undergo nonequilibrium dynamics, while use of the
Feshbach resonance ensures that $b$ atoms remain noninteracting and provide a
well-characterized reference \textquotedblleft beam\textquotedblright\ state.
Spatial separation between $a$ and $b$ atoms is required because it
effectively turns off their mutual interactions. After this point, we can
change the system dynamics to a \textquotedblleft nontrivial\textquotedblright%
\ Hamiltonian, as long spatial separation of $a$ and $b$ atoms is maintained.
For example, tunneling can be turned back on, the lattice geometry dynamically
transformed, or external fields applied.

The second beam splitter combines two operations. First, a Fourier transform
on atom spatial modes is applied, transforming site indices $i$ into
quasimomentum indices $k$. This is achieved by releasing atoms from the
lattice followed by (time-of-flight) expansion until all atoms have spatially
overlapping wave functions. Second, a $\pi/2$ pulse with tunable phase $\chi$
is applied to the internal states of all atoms, driving $|a\rangle
\rightarrow(e^{i\chi/2}|a\rangle+e^{-i\chi/2}|b\rangle)/\sqrt{2}$ and
$|b\rangle\rightarrow(-e^{i\chi/2}|a\rangle+e^{-i\chi/2}|b\rangle)/\sqrt{2}$.

Finally, we measure the observable ${S}_{k}=A_{k}^{\dagger}A_{k}%
-B_{k}^{\dagger}B_{k}$ by detecting the difference between the number of $a$
and $b$ atoms in quasimomentum $k$ $\emph{after}$ the second beam splitter.
Switching to the Heisenberg picture and using $A_{k}\rightarrow(e^{i\chi
/2}A{_{k}+e^{-i\chi/2}}B{_{k}})${$/\sqrt{2}$} and $B${$_{k}\rightarrow
(-e^{i\chi/2}A_{k}+e^{-i\chi/2}B_{k})/\sqrt{2}$ at the second beam splitter,
we find }that {the observable is equivalent to $S_{k}=$}$A{_{k}^{\dagger}%
}B{_{k}e^{-i\chi}+}B{_{k}^{\dagger}}A{_{k}e^{i\chi}}$ with expectation values
{taken with }respect to wave functions just \emph{before} the second beam splitter.

Because the many-body wave function is separable in the internal atomic states
before the $\pi/2$ pulse, it follows that $\langle S_{k}\rangle=\langle
A_{k}\rangle\langle B_{k}^{\dagger}\rangle{e^{-i\chi}}+\langle A_{k}^{\dagger
}\rangle\langle B_{k}\rangle{e^{i\chi}}$. The noninteracting reference atoms
give $\langle B_{k}\rangle=\beta g_{k}$ with $\beta=\sqrt{\bar{n}}e^{i\theta}$
and the complex-valued $g_{k}$ depends only on lattice geometry and
distribution of $b$ atoms. In particular, $g_{k=0}=\sqrt{M}/2$ for any lattice
geometry and atom distribution. We therefore find
\begin{equation}
\langle S_{k}\rangle=2\sqrt{\bar{n}}|g_{k}|\langle X_{k}\rangle, \label{Eq1}%
\end{equation}
where $X_{k}$ is the quadrature observable with tunable phase $\zeta_{k}%
=\arg(e^{i(\theta+\chi)}g_{k}).$ In contrast, regular collapse-and-revival
experiments \cite{Greiner2002,Will2010,Sebby2007,Anderlini2007}, using atoms
in only one spin state, measure the quasimomentum distribution $R_{k}%
=A_{k}^{\dagger}A_{k}$.

\section{Implementations}

\begin{figure}[ptb]
\begin{center}
\includegraphics[scale=1,trim=0 0 0 0,angle=0,clip]{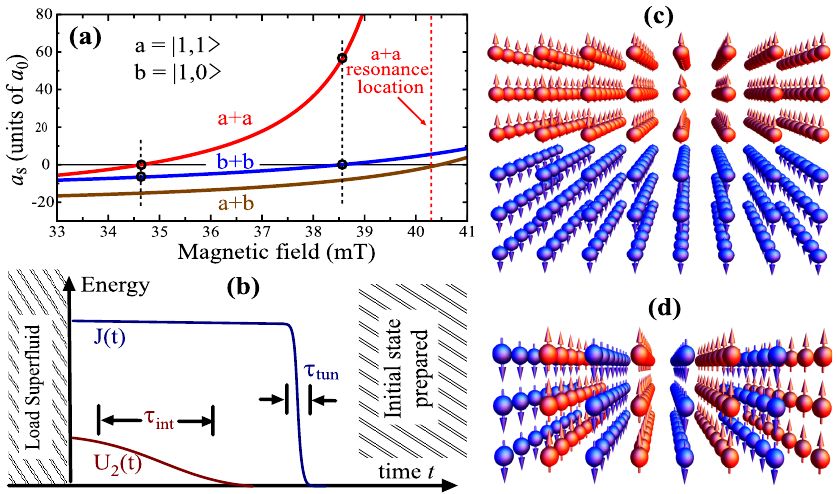} \vspace*{-17pt}
\end{center}
\caption{(Color online) (a) Scattering lengths $a_{s}$ of $^{39}$K collisions
versus magnetic field $B$ near a Feshbach resonance located at $B\approx40$
mT. Curves are labeled by $|fm\rangle=|11\rangle$ and $|10\rangle$, the
energetically lowest hyperfine channels of $^{39}$K. Two operating points for
the interferometer are shown by black-dotted lines (here $1a_{0}=0.0529$ nm).
(b) The interferometer initial state, a coherent state of noninteracting
atoms, is created starting from a superfluid of interacting atoms by using a
Feshbach resonance to slowly turn off the interaction strength $U_{bb}$,
followed by faster turn off of the tunneling rate $J$. See text for time and
energy scales. Panels (c) and (d) show two possible atom distributions after
the site-specific $\pi$ pulse. In panel (c) atoms in state $a$ (red spheres)
and $b$ (blue spheres) are located in the top and bottom half, respectively.
In panel (d) planes of state $a$ and $b$ atoms alternate.}%
\label{preparation}%
\end{figure}

Realization of the interferometer requires two internal atomic states that
have little or no two-body collisional loss, while allowing for tuning of
$b+b$ collisions to turn off interactions. We suggest the two energetically
lowest hyperfine states of bosonic alkali-metal atoms as only weak magnetic
dipole-dipole and second-order spin-orbit interactions can cause inelastic
loss. Candidates are $^{87}$Rb near the narrow resonance at a magnetic field
$B$ of 100.7 mT \cite{Marte2002} and $^{39}$K near a much broader resonance at
40.2 mT \cite{DErrico2007}, the latter requiring less stringent field control.
Other examples can be found in \cite{Chin2010}. Figure~\ref{preparation}(a)
shows the scattering length $a_{s}$ of the $^{39}$K resonance as a function of
$B$ and indicates two field values where one of the two hyperfine states has
zero scattering length and, hence, does not interact.

The initial state of the interferometer is prepared by starting with a
(superfluid) Bose condensate of \emph{interacting} $b$-state atoms in a weak
trap with level spacing $\hbar\omega_{\mathrm{w}}$, typically harmonic with
$\omega_{\mathrm{w}}/2\pi=1-100$ Hz; $\hbar$ is Planck's reduced constant.
After adiabatically turning on the optical lattice, the system can be
described by the single-band Bose-Hubbard Hamiltonian
\cite{Fisher1989,Freericks1994,Jaksch1998,vanOosten2001} with tunneling energy
$J$ and atom-atom interaction strength $U_{bb}$ proportional to $a_{s}$. In a
simple cubic lattice, a superfluid ground state requires $J\gg U_{bb}%
{/(5.83\times6)}$ at the end of the adiabatic ramp (the numerical factor
follows from a three-dimensional mean-field calculation). Typically,
$U_{bb}/h\sim1$ kHz away from Feshbach resonances. Due to atom-atom
interactions, at this stage the superfluid is only an approximate coherent state.

Figure~\ref{preparation}(b) shows the next step in initial state preparation.
Slowly turning off $U_{bb}$ at fixed $J$ using a Feshbach resonance over
timescale $\tau_{\mathrm{int}},$ the ground state approaches the desired
coherent state of noninteracting $b$ atoms. Choosing $\tau_{\mathrm{int}}%
\gg2\pi/\omega_{\mathrm{w}}$ minimizes excitation from zero quasimomentum.
After interactions are ramped-off, tunneling is suppressed by increasing
lattice depth with time scale $\tau_{\mathrm{tun}}$. Choosing $\tau
_{\mathrm{tun}}\ll2\pi/\omega_{\mathrm{bg}}$, where $\hbar\omega_{\mathrm{bg}%
}$ is the lattice band gap (typically $\omega_{\mathrm{bg}}/2\pi=10-100$ kHz),
suppresses excitation to higher bands.

The first beam splitter, a spatially selective $\pi$ pulse, can be
implemented, for example, by illuminating only the top half of the lattice
sites [see Fig.~\ref{preparation}(c)] with lasers driving optical Raman
transitions via off-resonant transitions to electronically excited states. The
interaction strength for $a$ atoms is $U_{aa}.$ This approach is limited by
the boundary over which the Raman laser intensities go to zero, which is no
sharper than the laser wavelength. Uncertainty in the fraction of $a$ versus
$b$ atoms leads to only small corrections of order $M^{-3/2}.$ An alternative
approach uses dynamically transformable lattices while simultaneously flipping
selected spin states. For example, Refs.~\cite{Anderlini2006,Anderlini2007}
used spin-dependent lattices to transform from a single- to double-well
lattice where the spin state of atoms alternates with lattice site. In
principle, alternating planes of $a$ and $b$ atoms can be created [see
Fig.~\ref{preparation}(d)] and our interferometric techniques used to study
correlations in two-dimensional many-body physics.

The second beam splitter involves release and expansion from the lattice
followed by a global $\pi/2$ pulse on the internal state of all atoms. The
latter operation requires no spatial selectivity and can be implemented with
Raman transitions, or microwave and radio-frequency radiation. The phases
$\pi/2$ and $\chi$ are controllable by choosing the polarization of the
radiation relative to the applied magnetic field; the required techniques have
been demonstrated in atomic clock and ion trap quantum computing
implementations \cite{Haffner2008}. Requirements on expansion time are
analyzed in \cite{Toth2008}, and we estimate that $\sim20$ ms is sufficient
for operation of the interferometer.

Finally, there are corrections due to interactions after atoms are released
from the lattice. Extending \cite{Toth2008}, we model the dominant effect of
\emph{coherent }collisions by a time-varying $U_{aa}$ proportional to the
inverse volume of the expanding local orbital. This leads to a small
additional evolution time of order $1/\omega_{\mathrm{bg}},$ typically
$\sim10$ $\mu$s, and independent of hold time $t$, and does not degrade the
performance of the interferometer.

Momentum changing, \emph{incoherent }collisions can also occur between atoms
originating from either the same or different lattice sites. We expect the
effect of these thermalizing collisions to be much smaller than those of
coherent collisions not only because of the rapidly decreasing density, but
also due to the rapidly decreasing relative collisional energies
($\approx\hbar\omega_{\mathrm{bg}}$ at the moment of release) during
expansion. The resulting lack of significant thermalization is supported by
existing collapse-and-revival experiments
\cite{Greiner2002,Will2010,Anderlini2006,Sebby2007}, which have demonstrated
the momentum distribution measurements needed for our proposed technique.

We also need to minimize effects from interactions between $a$ and $b$ atoms.
This can be achieved by delaying the $\pi/2$ pulse by $\gtrsim1/\omega
_{\mathrm{bg}}$ after release, but before significant phase shifts arise from
a combination of any magnetic field inhomogeneities and the differential
magnetic moment of the two spin states. Effects of field gradients after the
$\pi/2$ pulse should then only shift the relative location of the $k=0$ modes
for $a$- and $b$-atom clouds.

\section{Nonequilibrium dynamics}

The factorization property leading to Eq.~\ref{Eq1} holds even if tunneling
and other modified dynamics are induced for the interacting $a$ atoms after
the first beam splitter, making it possible to use the technique to measure
the quadrature of nonequilibrium fields for many types of Hamiltonians (e.g.
systems with spin-orbit couplings, different lattice geometries, or applied
fields). Below, we analyze the interferometer and predict the quadrature
evolution for the relatively simple evolution of the Bose-Hubbard model
without tunneling, i.e., $H_{\mathrm{MB}}=(U_{aa}/2)\sum_{i=1}^{M/2}%
a_{i}^{\dagger}a_{i}^{\dagger}a_{i}a_{i}$. Effective multibody interactions
\cite{Johnson2009,Johnson2012,Bissbort2012} are omitted for simplicity.

\begin{figure}[ptb]
\begin{center}
\includegraphics[scale=0.34,trim=0 0 0 0,angle=0,clip]{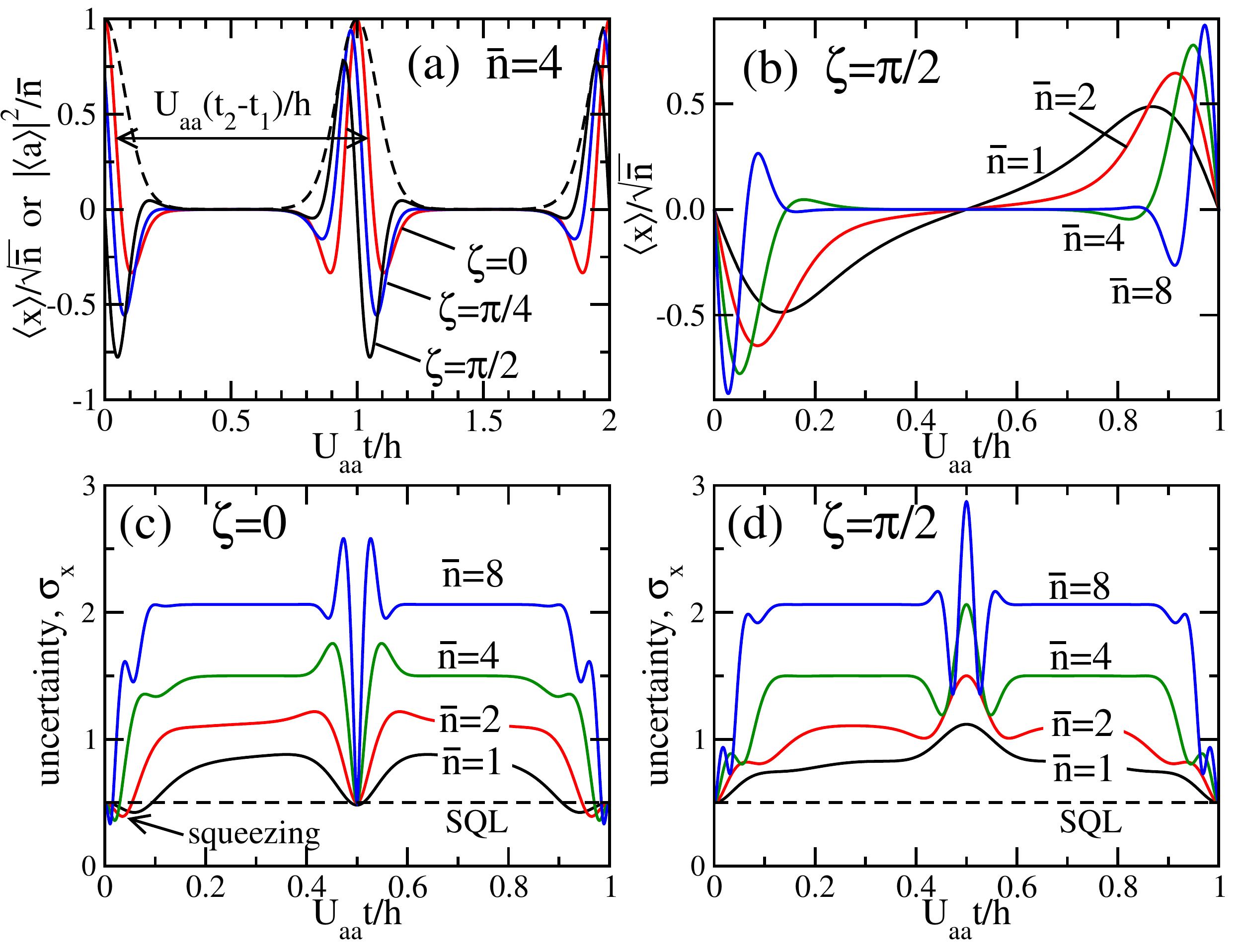}
\vspace*{-30pt}
\end{center}
\caption{(Color online) Panels (a) and (b) show the normalized single-site
quadrature $\langle x_{i}\rangle/\sqrt{\bar{n}}$ (solid lines) as a function
of evolution time in units of $h/U_{aa}$ for various quadrature phases and
mean atom number $\bar{n}$. The double-sided arrow indicates a possible time
interval for measuring $U_{aa}$. The density operator $|\langle a_{i}%
\rangle|^{2}/{\bar{n}}$ (dashed line) is only shown for $\bar{n}=4$. Panels
(c) and (d) show the quadrature uncertainties $\sigma_{xi}$ as a function of
evolution time and $\bar{n}$, for $\zeta_{0}=0$ and $\zeta_{0}=\pi/2,$
respectively. The dash-dotted line, labeled SQL, indicates $\sigma_{xi}=1/2$
for a coherent state.}%
\label{time}%
\end{figure}

We focus on the normalized $k=0$ quasimomentum observable $s_{0}(t)=\langle
S_{k=0}(t)\rangle/\bar{N}$, which is proportional to the order parameter.
Defining single-site quadrature operators $x_{i}=(e^{-i\zeta_{0}}%
a_{i}+e^{i\zeta_{0}}a_{i}^{\dagger})/2$, we find $s_{0}(t)=\langle
x_{i}(t)\rangle/(2\sqrt{\bar{n}}),$ after expanding the momentum operator in
terms of site operators and realizing that the time evolution is the same for
all sites. For $H_{\mathrm{MB}}$ we derive
\begin{equation}
\langle x_{i}(t)\rangle=\sqrt{\bar{n}}\exp([\cos(\phi)-1]{\bar{n}})\cos
({\bar{n}}\sin(\phi)+\zeta_{0}),
\end{equation}
with phase $\phi=U_{aa}t/\hbar$. Figures~\ref{time}(a) and \ref{time}(b) show
examples of the periodic evolution of $\langle x_{i}(t)\rangle/\sqrt{\bar{n}}$
for several values of $\zeta_{0}$ and $\bar{n}.$ The interferometer makes
possible direct experimental verification of these quadrature dynamics. The
time evolution of $\mathcal{R}_{k=0}/\bar{N}=|\langle a_{i}\rangle|^{2}%
/{\bar{n}}$ {(assuming large }${M}${),} determined by regular
collapse-and-revival measurements
\cite{Greiner2002,Anderlini2006,Sebby2007,Will2010}, is also shown.

The fluctuations in quadrature observables for nonequilibrium Bose gases in
deep lattices are also interesting. Figures~\ref{time}(c) and \ref{time}(d)
show the analytic time-dependent (single-site) uncertainty $\sigma
_{xi}=\langle(x_{i}-\langle x_{i}\rangle)^{2}\rangle^{1/2}$ as a function of
time. For a coherent state for any $\bar{n}$ and $\zeta_{0},$ $\sigma
_{xi}=1/2$, corresponding to the SQL and shown by horizontal dashed lines;
$\sigma_{xi}<1/2$ indicates quadrature squeezing. For most times and phases
$\zeta_{0},$ the uncertainty is $\sigma_{xi}>1/2$; in fact, a
\textquotedblleft plateau\textquotedblright\ develops for large $\bar{n}$
where $\sigma_{xi}$ approaches $\sqrt{2{\bar{n}}+1}/2$. For times near integer
multiples of $U_{aa}t/h$ and values of $\zeta_{0}$ near zero, however,
$\sigma_{xi}<1/2$, and the fluctuations in $\langle x_{i}\rangle$ are reduced
below the SQL. This is seen in Fig.~\ref{time}(c) for $\zeta_{0}=0,$ where we
find that the minimum value of $\sigma_{xi}$ approaches a constant
$\approx0.29$ for large $\bar{n}.$ In contrast, Fig.~\ref{time}(d) shows that
$\sigma_{xi}\ $for $\zeta_{0}=\pi/2$ is always antisqueezed.

\section{Interaction-based metrology}

In addition to directly measuring field quadratures, the interferometry can be
used to measure interaction strengths, although achieving optimal performance
is experimentally more demanding. The period $h/U_{aa},$ and hence $U_{aa},$
can be obtained by determining times $t_{1}$ and $t_{2}>t_{1}$ at which
$s_{0}(t)$ and its time derivative are the same. Figure~\ref{time}(a) shows an
example where $U_{aa}=h/(t_{2}-t_{1})$. Error propagation gives the fractional
uncertainty
\begin{equation}
\frac{\delta U_{aa}}{U_{aa}}=\frac{\sqrt{2}}{\pi\sqrt{M/2}}\left\vert \frac
{1}{d\langle x_{i}\rangle/d\phi}\right\vert \sqrt{\sigma_{xi}^{2}%
+\frac{\langle x_{i}\rangle^{2}}{2{\bar{n}}}}\,, \label{eq2}%
\end{equation}
with expectation values evaluated at phase $\phi_{1}=U_{aa}t_{1}/\hbar$,
assuming equal contributions at $t_{1}$ and $t_{2}$ added in quadrature. The
fractional uncertainty gives SQL scaling $\propto1/\sqrt{M/2}$ in number of
$a$-atom lattice sites, as expected for a site-separable wave function, and we
can view the interferometer as $M/2$ independent probes of $\bar{n}$
interacting atoms. The second term under the square root is due to the
uncertainty in total atom number $\bar{N}$. The uncertainty also scales as
$1/m$ in the number of oscillation periods $m$; for brevity we set $m=1.$

Minimizing $\delta U_{aa}/U_{aa}$ based on measurements of $x_{i}(t)$ involves
a trade-off between identifying optimal measurement times $t_{1}$, $t_{2}$ (or
phases $\phi_{1},\phi_{2}$) that maximize the slope $d\langle x_{i}%
\rangle/d\phi$ while minimizing $\sigma_{xi}.$ We find that the smallest
(optimal) fractional uncertainty
\begin{equation}
(\delta U_{aa}/U_{aa})_{\text{quad}}=(1/2\pi)M^{-1/2}\bar{n}^{-3/2},
\label{eq3}%
\end{equation}
is obtained from the quadrature observable with $\zeta_{0}=\pi/2$, and
measurements made when $\langle x_{i}\rangle=0$ and the slope $d\langle
x_{i}\rangle/d\phi$ is maximum. This occurs at phases $\phi_{1}=0$ and
$\phi_{2}=2\pi$, corresponding to times $t_{1}=0$ and $t_{2}=2\pi\hbar/U_{aa}%
$. Because $\langle x_{i}\rangle=0$ and $\sigma_{xi}=1/2$ at these times, the
result is independent of uncertainty in $\bar{N}$. For other $\zeta_{0},$ the
numerical prefactor in Eq.~(\ref{eq3}) is larger, but the scaling in $\bar{n}$
is the same. As noted in \emph{Implementations}, the small correction from
interactions during expansion can be included in the determination of
$t_{1,2}$.

At optimal measurement times the state is unsqueezed, and the super-Heisenberg
scaling $\bar{n}^{-3/2}$ derives from the nonlinear dependence of $d\langle
x_{i}\rangle/d\phi$ on $\bar{n}.$ Our quadrature interferometer yields the
best possible scaling behavior predicted for interaction-based metrology
schemes based on two-body interactions and nonentangled states
\cite{Boixo2008}. We have also analyzed the minimal uncertainty possible from
conventional collapse and revival measurements of $\langle\mathcal{R}%
_{k=0}(t)\rangle/\bar{N}$. We obtain the analytic estimate$\ (\delta
U_{aa}/U_{aa})_{\text{conv}}=(2/3)^{3/4}\pi^{-1}M^{-1/2}\bar{n}^{-3/4},$ and
have confirmed it is in good agreement with numerical evaluation. The scaling
is significantly less advantageous than what is possible using the quadrature
interferometry.

We can compare our predictions with conventional collapse-and-revival
experiments. Will \textit{et al.} \cite{Will2010} obtained $\delta
U_{aa}/U_{aa}\approx2\times10^{-2}$ with $\bar{N}=2\times10^{5}$ atoms and
${\bar{n}}=2.5$. For these parameters $(\delta U_{aa}/U_{aa})_{\text{conv}%
}=3\times10^{-4}.$ The larger experimental uncertainty is likely due to
fluctuations in the total atom number and lattice-laser intensities. For the
quadrature interferometer $(\delta U_{aa}/U_{aa})_{\text{quad}}=1.2\times
10^{-4}$ is, in principle, possible, giving nearly a factor of three
improvement even with ${\bar{n}}=2.5.$ As noted above, we also predict less
sensitivity to fluctuations in $\bar{N}$ using the quadrature technique.

\section{Conclusions}

Our analysis suggests that even for small (realistic) $\bar{n}$ it should be
possible to test predicted scaling behaviors for interaction-based metrologies
in an optical lattice system. There are many interesting directions for future
research. For example, are there interferometer states that give $\bar{n}%
^{-2}$ scaling per lattice site, as suggested by the analysis in
\cite{Boixo2008}? Can we design implementations where the important
information lies in $k\neq0$ quasi-momenta states? If so, this would be a
major application of our technique toward the study of nonequilibrium
many-body systems. Can we modify the interferometer to measure observables
such as $\langle a_{i}a_{j}\rangle,$ and use entanglement between atoms in
different sites to measure nonlocal field correlations? There are also
practical challenges facing implementations, including the effects of
inhomogeneities, lattice (laser) fluctuations, and limits on both atom number
per site and total atom number. Even with non-optimal measurements, however,
the interferometer technique developed here should expand the toolkit for
studying nonequilibrium dynamics in optical lattices.

\begin{acknowledgments}
P.R.J. and E.T. acknowledge support from the U.S. Army Research Office under
Contract No. 60661PH. P.R.J. also acknowledges computing resources provided by
the American University High Performance Computing System. E.T. acknowledges
support from a National Science Foundation Physics Frontier Center. Finally,
we thank Khan Mahmud, Lei Jiang, and Nathan Harshman for valuable discussions.
\end{acknowledgments}

\bibliography{refs}

\end{document}